\documentclass{amsart}      
\usepackage{amssymb}      
\usepackage{latexsym}      
      
\newcommand{\ZZ}{{\mathbb Z}}      
\newcommand{\RR}{{\mathbb R}}      
      
\newcommand{\NN}{{\mathbb N}}

\newtheorem{theorem}{Theorem}         
\newtheorem{remark}{Remark}         
\newtheorem{lemma}{Lemma}[section]         
\newtheorem{prop}[lemma]{Proposition}         
\newtheorem{coro}[lemma]{Corollary}         
\newtheorem{definition}[lemma]{Definition}         
\renewcommand\qedsymbol{$\Box$}      
      
\newcommand{\tr}{{\mathrm{tr}}}      
\newcounter{smalllist}

\newcommand{\CalA}{\mathcal{A} (\Omega,T)}  
\newcommand{\CalB}{\mathcal{B} }

\newcommand{\CalP}{\mathcal{P}}

\newcommand{\Oomega}{(\Omega,T)}  
\newcommand{\OOmega}{(\Omega,T)}  
\newcommand{\CalI}{\mathcal{I}}  
\newcommand{\CalD}{\mathcal{D}}  
\newcommand{\CalL}{\mathcal{L}}

\newcommand{\dist}{\mbox{dist}}

\begin{document}    
\title[Ergodic theory and the integrated density of states]
{An ergodic theorem for   
Delone dynamical systems and existence of the integrated density of states}  
  
\author[D. Lenz, P. Stollmann]{Daniel Lenz~$^{1,*}$,\    
Peter Stollmann~$^{2,*}$} \thanks{* Research partly supported    
by the DFG in the priority program Quasicrystals}

\begin{abstract}  We study strictly ergodic  Delone dynamical systems and    
prove an  ergodic theorem for Banach space valued functions on the   
associated  set of pattern classes.    
As an application, we  prove existence of the integrated density of states   
in the sense of uniform convergence in distribution for the associated   
random operators.    
\end{abstract}

\maketitle  
\vspace{0.3cm}  
\begin{center}  
\textit{Dedicated to J. Voigt on the occasion of his 60th birthday}  
\end{center}  
\vspace{0.3cm}  
\noindent  
$^{1}$ Fakult\"{a}t f\"{u}r     
Mathematik, Technische     
Universit\"{a}t Chemnitz,     
D-09107 Chemnitz, Germany;    
E-mail:  D.Lenz@mathematik.tu-chemnitz.de,\\[0.1cm]    
    
\noindent    
$^{2}$  Fakult\"{a}t f\"{u}r Mathematik, Technische Universit\"{a}t     
Chemnitz,     
D-09107 Chemnitz, Germany; E-mail:  P.Stollmann@mathematik.tu-chemnitz.de

\section{Introduction}\label{Introduction}  
This paper is concerned with Delone dynamical systems and the associated   
random operators.   
  
Delone dynamical systems can be seen  as the higher dimensional analogues   
of subshifts over finite alphabets. They have attracted particular attention   
as they can serve as models for so called quasicrystals. These are
substances,   discovered in 1984 by Shechtman, Blech, Gratias and Cahn \cite{She} (see 
the report \cite{Ni} of Ishimasa et al.  as well), which  exhibit features   
similar to crystals but are non-periodic. Thus, they belong to the reign of disordered   
solids and their distinctive feature is their special form of weak disorder.    
  
This form of disorder and its effects have been immensely studied in recent years, both    
from the theoretical and the   experimental point of view (see \cite{BM,Jan,Pat, Sen}   
and references therein). On the theoretical side, there  does not exist an axiomatic   
framework (yet) to describe quasicrystals. However, they are commonly modeled by either   
Delone dynamical systems or tiling dynamical systems \cite{Sen} (see \cite{l1,lp}   
for recent study of Delone sets as well). In fact, these two descriptions are essentially   
equivalent (see  e.g.   \cite{LS2}). The main focus of the theoretical
study lies  then   
on diffraction properties, ergodic and combinatorial features and the associated random   
operators (see  \cite{BM,Pat,Sen}).    
  
Here, we will deal with ergodic features of Delone dynamical systems and
the associated    random operators.   
The associated random operators  (Hamiltonians) describe basic quantum mechanical   
features of the models (e.g. conductance properties). In the one dimensional case,  starting   
with \cite{Cas,Sut},  various specific features of these Hamiltonians have been rigorously   
studied.  They include   purely singular continuous spectrum, Cantor spectrum and anomalous   
transport (see  \cite{Dam} for a recent review and an  extended bibliography). In the higher dimensional case our   
understanding is much more restricted. In fact, information on spectral types is completely   
missing. However, there  is  K-Theory providing some overall type information on possible gaps   
in the spectra. This topic was initiated  by Bellissard \cite{be2}  for almost periodic   
operators.  It has then been investigated for tilings starting with the work of  Kellendonk   
\cite{Kel} (see  \cite{BHZ,KP} for recent reviews).    
  
Now, our aim in this paper is to study the integrated density of states. This is a key   
quantity in the study of random operators. It gives some average type of information on the   
involved operators.   
  
We will show uniform existence of the density of states in the sense of uniform convergence   
in distribution of the underlying measures. This result is considerably stronger than the   
corresponding earlier results of Kellendonk \cite{Kel}, and Hof \cite{Hof1}, which only gave   
weak convergence. It fits well within the general point of view that quasicrystals should   
behave very uniformly due to their proximity to crystals.   
  
These results are particularly relevant as the limiting distribution may well have points of   
discontinuity. In fact, points of discontinuity are an immediate consequence of existence of   
locally supported eigenfunctions. Such eigenfunctions had already been observed in certain models   
\cite{ATF,FATK,KS,KF}. In fact, as discussed by the authors and Steffen Klassert  in \cite{KLS}, they can  easily be ``introduced'' without essentially   
changing the underlying Delone dynamical system.   
Moreover,  based on the methods  presented here, it is possible to show that points of   
discontinuity of the integrated density of states are exactly those energies for  which locally   
supported eigenfunctions exist (see  \cite{KLS} again).  
  
Let us emphasize that the limiting distribution is known to be
continuous for models on lattices \cite{DS} (and, in fact, even  stronger
continuity properties hold \cite{CS}). In these cases uniform
convergence of the distributions is an immediate    consequence of general measure theory.

To prove our result on uniform convergence (Theorem \ref{density}) we introduce a new   
method. It relies on studying convergence of   
averages in suitable  Banach spaces. Namely, the integrated density of states turns out to be   
given by an almost additive function with values in a certain Banach space (Theorem \ref{AAT}).    
To apply our method we prove  an ergodic theorem (Theorem \ref{BET}), for Banach space valued   
functions on the associated set of pattern classes.

This ergodic theorem may be of independent interest. It is an analog of a result of Geerse/Hof   
\cite{GH} for tilings associated to primitive substitutions. For real valued almost additive   
functions on linearly repetitive Delone sets related results  have   
been obtained by Lagarias and Pleasants \cite{lp}.  The   one-dimensional case has been studied   by one   
of the authors in \cite{Len2,Len3}.    
  
The  proof of our ergodic theorem  uses ideas  from the cited work of Geerse and Hof. Their work relies on suitable   
decompositions. These decompositions are naturally present in the framework of primitive   
substitutions. However, we need to construct them separately in the case we are dealing with.   
To do so, we use techniques of ``partitioning according to return words'' as introduced by   
Durand in \cite{Du1,Du2} for symbolic dynamics and later studied for tilings by Priebe    
\cite{Pri}. Note, however, that we need quite some extra effort, as we do not assume   
aperiodicity.   
  
\medskip  
  
The paper is organized as follows. In Section \ref{Setting} we introduce the notation and   
present our results. Section \ref{Decomposing} is devoted to a discussion of the relevant   
decomposition. The ergodic theorem is proved in Section \ref{Theergodic}.  Uniform convergence   
of the integrated density of states  is proven in Section \ref{Almost} after proving   
the necessary almost additivity.

\section{Setting  and results}\label{Setting}  
The aim of this section is to introduce some notation and to present our results, which   
cover part of what has been announced in \cite{LS1}.  In a companion paper \cite{LS2} more   
emphasis has been laid on the topological background and the basics of the groupoid   
construction and the noncommutative point of view.

\medskip  
  
For the remainder of the paper an integer $d\geq 1$ will be fixed and all Delone sets, patterns   
etc. will be subsets of $\RR^d$. The Euclidean norm on $\RR^d$ will be denoted by $\|\cdot\|$ as will the norms on various other normed spaces.   
For $s>0$ and $p\in \RR^d$, we let   
 $B(p,s)$ be the closed  ball  in $\RR^d$ around  $p$ with radius $s$.   
A subset $\omega$ of  $\RR^d$ is called  Delone set if there exist $r>0$ and $R>0$ such that  
\begin{itemize}  
  
\item $2r\leq \|x-y\|$ whenever $x,y\in \omega$ with $x\neq y$,  
  
\item $B(x,R)\cap \omega \neq \emptyset$ for all $x\in \RR^d$,   
\end{itemize}  
and the limiting values of $r$ and $R$ are called \textit{packing radius} and \textit{covering  
radius}, respectively.  
Such an $\omega$ will also be denoted as $(r,R)$-set.  Of particular interest will be the  
 restrictions of $\omega$ to bounded subsets of $\RR^d$. In order to treat these restrictions,   
we introduce the following definition.  
  
\begin{definition} (a) A pair  $(\Lambda,Q)$ consisting of a bounded  subset $Q$ of $\RR^d$   
and  $\Lambda\subset  Q$ finite is called  \textrm{pattern}. The set $Q$ is called the \textrm{support of the   
pattern}. \\  
(b) A pattern $(\Lambda,Q)$ is called \textrm{ball pattern} if  $Q=B(x,s)$  with   $x\in \Lambda$ for   
suitable $x\in \RR^d$ and $s>0$.   
\end{definition}  
  
The diameter and the volume of a pattern are defined to be the diameter and the volume of its   
support respectively. For patterns $X_1=(\Lambda_1,Q_1) $ and $X_2= (\Lambda_2,Q_2)$, we define   
$\sharp_{X_1} X_2$, the number of occurrences of  $X_1$ in $X_2$, to be the number of elements   
in $\{t\in \RR^d :  \Lambda_1 +t = \Lambda_2\cap(Q_1+t), Q_1 +t = Q_2\}$.  Moreover, for patterns   
$X_i=(\Lambda_i, Q_i)$, $i=1, \ldots, k$, and $X=(\Lambda,Q)$, we write $X=\oplus_{i=1}^k X_i$   
if $\Lambda=\cup \Lambda_i$, $Q=\cup Q_i$ and the $Q_i$ are disjoint up to their boundaries.   
  
For further investigation we will have to identify patterns which are equal up to translation.   
Thus, on the set of patterns we introduce an equivalence relation  by setting   
$(\Lambda_1,Q_1)\simeq (\Lambda_2, Q_2)$ if and only if there exists a $t\in \RR^d$ with   
$\Lambda_1 = \Lambda_2 + t$ and $Q_1=Q_2 + t$. The class of a pattern $(\Lambda,Q)$ is denoted   
by $[(\Lambda,Q)]$.  The notions of diameter, volume occurrence etc. can easily be carried over   
from patterns to pattern classes.

Every Delone set $\omega$ gives rise to a set of pattern classes,   
$\CalP (\omega)$ viz $\CalP (\omega)=\{ Q\wedge \omega : Q\subset\RR^d\: \mbox{bounded and   
measurable} \}$, and to a set of ball pattern classes $\CalP_B (\omega)) =\{ [B(p,s)\wedge   
\omega] : p\in \omega, s\in \RR\}$.  Here we set   
\begin{equation}  
Q\wedge \omega= (\omega \cap Q, Q).  
\end{equation}  
Furthermore, for arbitrary ball patterns $P$, we define $s(P)$ to be the radius of the   
underlying ball, i.e.  
\begin{equation}  
s(P) = s \;\:\mbox{for $P=[(\Lambda, B(p,s))]$}.  
\end{equation}  
  
For $s\in (0,\infty)$, we denote by $\CalP_B^s (\omega)$ the set of ball patterns with radius   
$s$.   
A Delone set is said to be of finite type if for every radius $s$ the set $\CalP_B^s (\omega)$   
is finite.  
  
The Hausdorff metric on the set of compact subsets of $\RR^d$ induces the so called \textit{natural   
topology} on the set of closed subsets of $\RR^d$. It is described in detail in \cite{LS2}  and   
shares some nice properties: Firstly, the set of all closed subsets of $\RR^d$ is compact in   
the  natural topology. Secondly, the natural action $T$ of $\RR^d$ on the closed sets    
given by  $T_t C\equiv C + t$ is continuous.

\begin{definition}  
(a) If $\Omega$  is a set of Delone sets that is invariant under the shift $T$ and  closed   
under the natural topology, then $\OOmega$ is  called a \emph{Delone dynamical system}  and   
abbreviated as \emph{DDS}.  \\  
(b) A DDS $\Oomega$  is said to be of \emph{finite local complexity} if $\cup_{\omega\in \Omega}   
P_B^s (\omega)$   
is finite for every $s>0$. \\  
(c) A DDS  $\OOmega$ is called an $(r,R)$-\emph{system}  if every $\omega \in \Omega$ is an   
$(r,R)$-set.\\  
(d) The set $\CalP (\Omega)$ of patterns classes associated to a DDS $\Omega$ is defined   
by $\CalP (\Omega)=\cup_{\omega\in \Omega} \CalP (\omega)$.   
 \end{definition}  
  
Due to compactness of the set of all closed sets in the natural topology a DDS $\Omega$ is compact.   
  
Let us record the following notions of ergodic theory along with an equivalent ``combinatorial''   
characterization available for Delone dynamical systems (see e.g. \cite{lp, Sol1} for further discussion and references) : $\OOmega$ is called \textit{aperiodic}   
if $T_t \omega \neq \omega$ whenever $\omega \in \Omega$ and $t\in \RR^d$ with $t\neq 0$.   
Is is called \textit{minimal} if every orbit is dense. This is equivalent to $\CalP (\Omega)=   
\CalP(\omega)$ for every $\omega \in \Omega$. This latter property is called \textit{local   
isomorphism property} in the tiling framework \cite{Sol1}. It is also referred to as   
\textit{repetitivity}. Namely, it is equivalent to the existence of an $R(P)>0$ for every $P\in   
\CalP(\Omega)$ such that $B(p,R(P))\wedge \omega$ contains a copy of $P$ for every $p\in \RR^d$   
and every $\omega\in \Omega$.  Note  also that every minimal DDS is an $(r,R)$-system.

We are interested in ergodic averages. More precisely,  we will take means of suitable   
functions along suitable sequences of patterns  and pattern classes. These functions and   
sequences will be introduced next.  Here and in the sequel we will use the following notation:   
For $Q\subset \RR^d$and $h>0$ we define   
$$Q_h \equiv \{ x\in Q : \dist(x,\partial Q) \geq h\}, \;\:  Q^h\equiv \{ x\in \RR^d:   
\dist(x,Q)\leq h\},$$  
where, of course, $\dist$ denotes the usual distance and $\partial Q$ is the boundary of $Q$.    
Moreover, we denote the Lebesgue measure of a measurable subset $Q\subset \RR^d$ by $|Q|$.    
Then, a sequence $(Q_n)$ of subsets in $\RR^d$ is called a \textit{van Hove sequence} if the   
sequence   
$(|Q_n|^{-1} |Q_n^h \setminus Q_{n,h}|) $ tends to zero for every $h\in (0,\infty)$. Similarly, a   
sequence  $(P_n)$  of pattern classes, (i.e.  $P_n = [(\Lambda_n,Q_n)]$ with suitable   
$Q_n,\Lambda_n$) is called a \textit{van Hove sequence} if $Q_n$ is a van Hove sequence.   
(This is obviously well defined.)  
We can now discuss unique ergodicity. A dynamical system $\OOmega$ is called   
\textit{uniquely ergodic} if it admits only one $T$-invariant measure (up to normalization).   
For a Delone dynamical system, this is equivalent to the fact that for every pattern class $P$   
the frequency  
\begin{equation}\label{ue}  
f(P)\equiv \lim_{n\to \infty} |Q_n|^{-1} \sharp_P (\omega\wedge Q_n),  
\end{equation}  
exists uniformly in $\omega\in \Omega$ for every van Hove sequence $(Q_n)$. This   
equivalence was shown in Theorem 1.6 in \cite{LS2} (see \cite{LMS} as well). It goes   
back to \cite{Sol1}, Theorem 3.3,  in the tiling setting.

\begin{definition} \label{definitionAlAd}Let $\Omega$ be a DDS and  $\CalB$ be a vector space   
with seminorm $ \|\cdot\|$. A function $F : \CalP(\Omega) :  \longrightarrow \CalB$ is   
called \emph{almost additive (with respect to $\|\cdot\|$)} if there exists  a function   
$b : \CalP(\Omega)\longrightarrow [0,\infty)$ (called \emph{associated error  
function})  and a constant $D>0$  such that  
\begin{itemize}  
\item[(A1)] $\| F( \oplus_{i=1}^k P_i) - \sum_{i=1}^k F(P_i) \| \leq \sum_{i=1}^k b(P_i)$,  
\item[(A2)] $\|F(P)\| \leq D |P| + b(P)$.  
\item[(A3)] $b(P_1) \leq b(P) + b(P_2)$ whenever $P=P_1\oplus P_2$,  
\item[(A4)] $\lim_{n\to \infty} |P_n|^{-1} b(P_n)=0$   
for every van Hove sequence $(P_n)$.   
\end{itemize}

\end{definition}

Now, our first  result reads as follows.  
  
\begin{theorem}\label{BET} For a minimal, aperiodic DDSF $\Oomega$ the following are equivalent:\\  
(i) $\Oomega$ is uniquely ergodic.\\  
(ii) The limit $lim_{k\to \infty}  |P_k|^{-1}  F(P_k)$ exists for every van Hove sequence   
$(P_k)$ and every almost additive $F$  on $\Oomega$ with values in a Banach space.  
  
\end{theorem}  
  

The proof of the the theorem  makes use of completeness of the Banach space in crucial manner.   
However, it does not use the nondegeneracy of the norm. Thus, we get the following corollary   
(of its proof).  
  
\begin{coro}\label{lc} Let $\OOmega$ be aperiodic and strictly ergodic.  Let the vector space   
$\CalB$ be complete with respect to the topology induced by the  seminorms $\|\cdot\|_\iota$,    
$ \iota \in \CalI$. If $F :\CalP\longrightarrow \CalB$  is  almost additive with respect to   
every $\|\cdot\|_\iota$. $\iota\in \CalI$ , then $\lim_{k\to \infty} |P_k|^{-1} F(P_k)$ exists   
for every van Hove sequence $(P_k)$ in $\CalP(\Omega)$.   
\end{coro}  
  
The main theorem may also be rephrased as a result on additive functions on Borel sets. As   
this may also be of interest we include a short discussion  
  
\begin{definition} Let $\OOmega$ be a DDS and $\CalB$ be a Banach space.  Let $\mathcal{S}$ be   
the family of bounded measurable sets on $\RR^d$. A function $F : \mathcal{S}\times \Omega   
\longrightarrow \CalB$ is called almost additive if there exists a function $b : \mathcal{S}   
\longrightarrow [0,\infty)$  and  $D>0$ such that  
  
\begin{itemize}  
  
\item[(A0)] $b(Q) = b(Q+t)$ for arbitrary $Q\in \mathcal{S} $ and $t\in\RR^d$  and   
$\| F_\omega (Q) - F_\omega (Q')\| \leq b(Q)$ whenever $\omega \wedge Q = \omega \wedge Q'$.  
\item[(A1)] $\| F_\omega (\cup_{j=1}^n Q_j) - \sum_{j=1}^n F_\omega (Q_j)\| \leq \sum_{j=1}^n   
b (Q_j)$ for arbitrary $\omega\in \Omega$ and $Q_j\in \mathcal{S}$ which are disjoint up to   
their boundaries,  
\item[(A2)] $\|F_\omega(Q)\| \leq D |Q| + b(Q)$.  
\item[(A3)] $b(Q_1) \leq b(Q) + b(Q_2)$ whenever $Q=Q_1\cup Q_2$ with $Q_1$ and $Q_2$ disjoint   
up to their boundaries.    
\item[(A4)]$\lim_{k\to \infty} |Q_k|^{-1} b(Q_k)$ for every van   
Hove sequence $(Q_k)$ .  
  
\end{itemize}

\end{definition}

\begin{coro}\label{mengenversion} Let $\OOmega$ be  a strictly ergodic DDS and $F : \mathcal{S}\times \Omega$ be   
almost additive. Then $\lim_{k\to \infty} |Q_k|^{-1} F_\omega (Q_k)$ exists for arbitrary   
$\omega\in \Omega$ and every van Hove sequence $(Q_k)$ in $\RR^d$ and the convergence is   
uniform on $\Omega$.   
  
\end{coro}

Our further results concern selfadjoint operators in a certain  $C^\ast$ algebra associated   
to $\Oomega$. The construction of this $C^\ast$ algebra has been given in our earlier work   
\cite{LS1,LS2}. We recall the necessary details next.

\begin{definition} Let $\Oomega$ be a DDSF.  A family  $(A_\omega)$ of bounded  operators   
$A_\omega : \ell^2 (\omega) \longrightarrow \ell^2 (\omega)$ is called a \emph{random operator of   
finite range} if there exists a constant $s>0$ with  
\begin{itemize}  
\item $A_\omega (x,y) =0$ whenever $\|x-y\| \geq s$.  
\item  $A_\omega (x,y)$  only depends on the pattern class   
of $((K(x,s)\cup K(y,s))\wedge \omega)$ .   
\end{itemize}  
The smallest such $s$ will be denoted by $R^A$.   
\end{definition}  
The operators of finite range form  a  $*$-algebra under the obvious operations.   
There is a natural $C^\ast$-norm on this algebra and its completion is a $C^\ast$-algebra   
denoted as $\CalA$ (see \cite{BHZ,LS1,LS2} for details). It consists again of families   
$(A_\omega)_{\omega\in \Omega}$ of operators $A_\omega : \ell^2 (\omega) \longrightarrow   
\ell^2 (\omega)$.   
  
Note that for selfadjoint $A\in \CalA$ and bounded $Q\subset\RR^d$     
the restriction $A_\omega|_Q$ defined on $\ell^2(Q\cap\omega)$ has finite     
rank.     
Therefore, the spectral counting function    
$$    
n(A_\omega,Q)(E):=\#\{ \mbox{ eigenvalues of }A_\omega|_Q\mbox{  below  }E\}    
$$    
is finite and    
$\frac{1}{|Q|}n(A_\omega,Q)$ is the distribution function of the measure    
$\rho(A_\omega,Q)$, defined by    
$$    
\langle \rho(A_\omega,Q),\varphi\rangle:=     
\frac{1}{|Q|}\mbox{tr}(\varphi(A_\omega|_Q))\mbox{  for  }\varphi\in C_b(\RR). $$   
  
These spectral counting functions are obviously elements of the vector space  $\CalD$    
consisting of all bounded right continuous functions $f : \RR\longrightarrow \RR$  for   
which $\lim_{x\to -\infty} f(x) = 0$ and $\lim_{x\to \infty} f(x)$ exists.  Equipped with the    
supremum norm $\|f\|_\infty\equiv \sup_{x\in \RR} |f(x)|$ this vector space is a Banach space.   
It turns out that the spectral counting function is essentially an almost additive function.   
More precisely the following holds.   
  
\begin{theorem}\label{AAT} Let $\OOmega$ be a DDS. Let $A$ be an operator of finite range.   
Then $F^A : \CalP (\Omega) \longrightarrow \CalD$, defined by $F^A (P) \equiv n(A_\omega, Q_{R^A})$  
 for $P=[(\omega\wedge Q)]$ is a well defined  almost additive function.   
\end{theorem}  
\begin{remark}{\rm The theorem seems to be new even in  the one-dimensional case. (There, of   
course, it is very easy to prove.)}  
\end{remark}

Based on the foregoing two theorems it is rather clear  how to show existence of the limit   
$\lim_{k\to \infty} |Q_k|^{-1} n(A_\omega,Q_n)$ for van Hove sequences $(Q_k)$. This limit is   
called the spectral density of $A$. It is possible to express this limit in closed form using   
a certain trace on a von Neumann algebra \cite{LS1,LS2}. We will not discuss this trace here,   
but rather directly give a closed expression. This will be done next.   
To each selfadjoint element $A\in \CalA$, we associate the measure $\rho^A$ defined on $\RR$ by  
$$\rho^A (F) \equiv \int_\Omega \tr_\omega (M_f (\omega) \pi_\omega( F(A))) d\mu(\omega).$$  
Here, $\tr_\omega$ is the standard trace on the bounded operators on $\ell^2 (\omega)$,  $f$ is   
an arbitrary nonnegative  continuous function with compact support on $\RR^d$ with $  
\int_{\RR^d} f(t) dt=1$ and $M_f (\omega)$ denotes the operator of multiplication with $f$ in   
$\ell^2 (\omega)$ (see  \cite{LS1,LS2} for details). It turns out that $\rho^A$ is a spectral   
measure for $A$ \cite{LS2}.  Our result on convergence of the integrated density of states is   
the following.

\begin{theorem}\label{density} Let $\Oomega$ be a strictly ergodic DDSF.  
 Let $A$ be a selfadjoint operator of finite range and $(Q_k)$ be an arbitrary van Hove   
sequence. Then  the distributions $E\mapsto  \rho_{Q_k}^{A_\omega} ((-\infty, E])$ converge     
to the distribution $E\mapsto \rho^A((-\infty,E])$  with respect to $\|\cdot\|_\infty$  and   
this convergence is uniform in $\omega\in \Omega$.   
\end{theorem}  
\begin{remark}{\rm (a)  The usual  proofs of existence of the integrated density of states only   
yield weak convergence of the measures.\\   
(b) The proof of the  theorem uses   the fact, already established in \cite{LS2,LS4}, that the   
measures $\rho(A_\omega, Q_n)$ converge weakly towards the measure $\rho^A$ for every   
$\omega\in \Omega$ and $A\in \CalA$.}   
\end{remark}

As mentioned in the preceeding remark, the usual proofs of existence of the integrated density of states   
only give weak convergence of the measures $\rho_{Q_n}^{A_\omega}$.  Weak  convergence of   
measures does in general not imply convergence in distribution. Convergence in distribution   
will follow, however, from weak convergence if the limiting distribution is continuous.     
Thus, the theorem is particularly interesting   in view of the fact that the limiting   
distribution can have points of discontinuity.     
  
Existence of such discontinuities  is rather remarkable as it is completely different from the   
behaviour of random operators associated to models with higher disorder.  It turns out that a    
very precise understanding of this phenomenon can be obtained invoking the results presented   
above.  Details of this will be given separately \cite{KLS}. Here, we only mention the following   
theorem.  
  
\begin{theorem} Let $\Oomega$ be a strictly ergodic DDSF and $A$ an operator of finite range   
on $\Oomega$. Then, $E$ is a point of discontinuity of $\rho^A$ if and only if there exists a   
locally supported eigenfunction of $A_\omega - E $ for one (every) $\omega \in \Omega$.   
\end{theorem}

\section{Decomposing Delone sets} \label{Decomposing}  
This section provides the main geometric ideas underlying the proof of our
ergodic theorem, Theorem \ref{BET}. We  first discuss how to decompose a given Delone set in
finite pieces, called cells,  in a   
natural manner, Proposition \ref{crucialcell}. This is based on the
Voronoi construction, as given in \eqref{voronoidef} and Lemma  \ref{voronoi},  together with a certain way   
to obtain Delone sets from a given Delone set and a pattern. This
decomposition will be done on an increasing sequence of scales. As
mentioned already, here we  use ideas from \cite{Du1,Pri}. Having described
these decompositions,  our main concern
is  to study  van Hove type properties of the induced sequences of
cells. This study will be undertaken in a series of lemmas yielding  as
main results Proposition \ref{crucialvH} and Proposition
\ref{crucialvHstern}. Here, the proof of Proposition
\ref{crucialvHstern} requires quite some extra effort (compared with the 
proof of Proposition \ref{crucialvH}) as we have to cope with periods.

\medskip  
  
We start with a discussion  of the well known Voronoi construction.  Let $\omega$ be an   
$(r,R)$-set. To an arbitrary $x\in \omega$ we associate the  Voronoi cell $V(x,\omega)  
\subset \RR^d$ defined by  
\begin{eqnarray} \label{voronoidef}  
 V(x,\omega)&\equiv& \{p\in \RR^d : \|p-x\|\leq \|p-y\|\;\: \mbox{for all $y\in \omega$ with   
$y\neq x$}\}\\  
&=& \cap_{y\in \omega,y\neq x} \{p\in \RR^d :  \| p -x \|\leq \| p -y\|\}.  
\end{eqnarray}  
Note that $\{p\in \RR^d : \| p -x \|\leq \| p -y\|\}$ is a half-space. Thus, $V(x,\omega)$   
is a convex set. Moreover, it is obviously closed and bounded and therefore compact. It turns   
out that $V(x,\omega)$ is already determined by the elements of $\omega$ close to $x$. More   
precisely, the following is valid.   
  
\begin{lemma} \label{voronoi} Let $\omega$ be an $(r,R)$-set. Then,  
$V(x,\omega)$ is determined by  $B(x,2R)\wedge \omega$, viz  
$V(x,\omega)\equiv \cap_{y\in B(x,2R)} \{p\in \RR^d : \| p -x \|\leq \|
p -y\|\}$. Moreover,  $V(x,\omega)$ is contained in $B(x,R)$.   
\end{lemma}   
{\it Proof.} The first statement follows from Corollary 5.2 in
\cite{Sen} and the second one  is a consequence of   
Proposition 5.2 in \cite{Sen}. \hfill \qedsymbol  
  
\medskip  
  
Next we describe our notion of derived Delone sets. Let $\omega$ be an $(r,R)$-set and $P$   
be a ball pattern class with $P\in \CalP(\omega)$. Then, we define the
Delone set derived from $\omega$ by $P$, denoted as $\omega_P$,  to be the set   of all occurrences of $P$  in $\RR^d$, i.e.  
$$\omega_P \equiv \{t\in \RR^d : [B(t,s(P))\wedge \omega] = P\}.$$  

Now, let $\OOmega$ be minimal.  Choose $\omega \in \Omega$ and $P\in \CalP_B(\Omega)$. Then, the   
Voronoi construction applied to $\omega_P$ yields a decomposition of $\omega$ into cells  
$$C(x,\omega,P)\equiv V(x,\omega_P)\wedge \omega, \;\:x\in \omega_P.$$  
More precisely,   
$$\RR^d = \cup_{x\in \omega_P} V(x,\omega_P), \;\:\mbox{and}\;\: int(V(x,\omega_P))\cap   
int(V(y, \omega_P))=\emptyset,$$  
whenever $x\neq y$. Here, $int(V)$ denotes the interior of $V$.  This way of decomposing   
$\omega$ will be called the $P$-decomposition of $\omega$.  It is a crucial fact that each   
$C(x,\omega,P)$ is  already determined by  
$$ B(x,2 R(P))\wedge \omega, $$  
as can be seen  by Lemma \ref{voronoi}, where $R(P)$ denotes the covering radius of $\omega_P$.  Thus, in particular the following holds.   
  
\begin{prop}\label{crucialcell} Let $\OOmega$ be a minimal DDS and  $P\in \CalP_B(\Omega)$ be   
fixed.  Let $\omega\in\Omega$ with $0\in \omega_P$ and set $Q =B(0,2 R(P))\wedge  \omega$.   
Then $C(Q) \equiv [V(0,\omega_P)\wedge \omega]$ depends only on $[Q]$ (and not on $\omega$).   
Moreover, if $\widetilde{C}$ is a cell occurring in the $P$-decomposition of some $\omega_1\in   
\Omega$, then $[\widetilde{C}]= C(Q)$ for a suitable $\omega\in\Omega$ with $0\in \omega_P$.   
\end{prop}  
  
The proposition says that the occurrences of certain cells in the $P$-decompositions are   
determined by the occurrences of the larger $Q\in \{ [B(x,2 R(P))\wedge  \omega] :   
x\in \omega_P, \omega \in \Omega\}$. The proposition does not say that different $Q$   
induce different $C(Q)$ (and this will in fact not be true in general).

\medskip

The main aim is now to study the decompositions associated to an
increasing sequence of ball pattern classes $(P_n)$. We begin by  studying minimal and maximal distances between occurrences of a ball pattern class $P$.   
We need  the following  definition.  
  
\begin{definition} Let  $\OOmega$ be  minimal and $P\in \CalP_B$ be arbitrary.  
Define $r(P)$ as the packing radius of $\omega_P$, i.e., by  
$ r(P)\equiv \frac12\inf\{ \|x-y\| : x\neq y, x,y\in \omega_P, \omega \in \Omega\}$ and the   
occurrence radius $R(P)$ by  
$R(P)\equiv \inf\{ R>0 :  \sharp_P ( [B(p,R)\wedge \omega])\geq 1 \;\:\mbox{for every   
$p\in \RR^d$ and $\omega \in \Omega$}\}.$  
\end{definition}

\begin{lemma}  \label{anfang}  Let  $\OOmega$ be minimal. Then   
$R(P)\equiv \min\{ R>0 :  \sharp_P ( [B(p,R)\wedge \omega])\geq 1 \;\:\mbox{for every $p\in \RR^d$   
and $\omega \in \Omega$}\}.$ Moreover,   
$\omega_P$ is an $(r(P),R(P))$-set for every $\omega \in \Omega$.   
\end{lemma}  
{\it Proof.}   
We show that the infimum is a minimum. Assume the contrary and set $R':=R(P)$. Then there   
exist $p\in \RR^d$ and $\omega \in \Omega$ such that $B(p,R')\wedge \omega$ does not contain   
a copy of $P$. However, by definition of $R'$, $B(p,R'+\epsilon)\wedge \omega$ contains a copy   
of $P$ for every $\epsilon>0$. As $\omega$ is a Delone set, $B(p,R'+1)\wedge \omega$  contains   
only finitely many copies of $P$  and a contradiction follows.  The last statement of the lemma   
is immediate. \hfill\qedsymbol   
  
\medskip  
  
We will have to deal with Delone sets which are not aperiodic. To do so the following notions   
will be useful.  For a minimal DDS $\OOmega$ let  $\CalL\equiv \CalL(\Omega)$ be the periodicity lattice   
of $\Oomega$, i.e.   
$$\CalL\equiv \CalL(\Omega)\equiv \{t\in \RR^d : T_t \omega = \omega \;\:\mbox{for all $\omega \in \Omega$}\}.   
$$  
Clearly, $\CalL$ is a subgroup of $\RR^d$; it is discrete, since every $\omega$ is discrete.  
Thus, (see e.g.  Proposition 2.3 in \cite{Sen}), $\CalL$ is a lattice in $\RR^d$, i.e. there  exists $D(\CalL) \in \NN$ and vectors    
$e_1,\ldots, e_{D(\CalL)}\in \RR^d$ which are linearly independent (in
$\RR^d$)  such that 

$$\CalL= \mbox{Lin}_{\ZZ}\{ e_j : j=1,\ldots D(\CalL)\} \equiv \{\sum_{j=1}^  
{D(\CalL)} a_j e_j : a_j\in \ZZ, j=1, \ldots, D(\CalL)\}.$$  
We define $r(\CalL)$ by   
\begin{equation*}
r(\CalL)  \equiv \left\{\begin{array}{r@{\quad:\quad}l}
  \infty    &;\: \mbox{if}\; \CalL=\{0\}\\  
\frac12\min\{\|t\| : t\in \CalL\setminus\{0{}\} & ;\:\mbox{otherwise}. 
\end{array}\right.
\end{equation*}

Next, we provide  a result on minimal distances, viz. Lemma \ref{dist}. Variants of this result have been given in the   
literature on tilings \cite{Pri} and on symbolic dynamics \cite{Du1}. To 
prove it in our context we recall   
the following lemma concerning the natural topology from \cite{LS5}:  
  
\begin{lemma} \label{conv}  
  A sequence $(\omega_n)$ of Delone sets converges to $\omega\in\CalD$  
  in the natural topology if and only if there exists for any $l>0$ an  
  $L>l$ such that the $\omega_n\cap B(0,L)$ converge to $\omega\cap  
 B(0,L)$ with respect to the Hausdorff distance as $n\to\infty$.  
\end{lemma}  
  
\begin{lemma}\label{dist}  
 Let $\OOmega$ be  minimal. Let $(P_n)$ be a sequence of ball pattern classes with $s(P_n)  
\longrightarrow \infty$, $n\longrightarrow \infty$. Then, $$\liminf_{n\to \infty}  r(P_n)    
\geq r(\CalL).$$  
\end{lemma}  
{\it Proof.} As $\Oomega$ is minimal, it is an $(r,R)$-system.  Assume that the claim is false.   
Thus, there exists a sequence $(P_n)$ in $\CalP_B(\Omega)$ with $s(P_n)\to \infty$,   
$n\to \infty$, but $r(P_n) \leq C$ with a suitable constant $C>0$ with $C< r(\CalL)$. Then,   
there exist $\omega_n\in \Omega$ and $t_n\in \RR^d$ with $\|t_n\|\leq \frac12 C$ (and, of course,   
$\|t_n\|\geq \frac12 r$) with  
\begin{equation}\label{localperiodic}  
B(0,s(P_n))\wedge \omega_n = B(0,s(P_n)) \wedge (\omega_n -t_n).  
\end{equation}  
By compactness of $\Omega$ and $B(0,\frac12 C)$, we can assume without loss of generality   
that $\omega_n\to \omega$ and $t_n\to t$, with $t\in B(0,C)$,  $n\to \infty$. Thus,   
\eqref{localperiodic} implies    
\begin{equation}\label{period} \omega = \omega -t.
\end{equation}
In fact, let $p\in\omega$. Fix $R>0$ such that $p\in\omega\cap B(0,R)$. By Lemma  
\ref{conv} we find $p_n\in\omega_n\cap B(0,R)$, for $n$ sufficiently large, such that $p_n\to p$  
for $n\to\infty$. Assuming $R<s(P_n)$ and utilizing \eqref{localperiodic} we find $q_n\in  
\omega_n$ such that $p_n=q_n-t_n$. Since $q_n\to p+t$ and $\omega_n\to\omega$ we see that  
$q=p+t\in\omega$ leaving us with  
$$  
\omega\cap B(0,R)\subset (\omega -t)\cap B(0,R) .  
$$   
By symmetry and since $R$ was arbitrary, this gives \eqref{period}.  
Minimality yields that \eqref{period} extends to all $\omega\in\Omega$.  
As $0<r\leq \|t\|\leq C < r(\CalL)$ , this gives a contradiction. \hfill\qedsymbol   
  
\medskip  
  
\begin{definition} For a compact convex set $C\subset \RR^d$ denote by  $s(C) >0$  the   
\emph{inradius} of $C$, i.e. the largest $s$ such that $C$ contains a ball of radius $s$.   
\end{definition}

In the sequel we write $\omega_{n,P_n}:=(\omega_n)_{P_n}$ to shorten notation 

\begin{lemma}\label{VvH} Let $(\Omega_n,T)$, $n\in \NN$ be a family of minimal DDS. Let    
a pattern class $P_n\in \CalP(\Omega_n)$,  an $\omega_n\in \Omega_n$ and $x_n\in \omega$    
be given for any $n\in \NN$.  If $r(P_n)\longrightarrow \infty$, $n\longrightarrow \infty$,   
then  $s(V(x_n, \omega_{n,P_n}))\to \infty$, for $n\to  \infty$.
\end{lemma}   
{\it Proof.} Without loss of generality we can assume that $x_n=0$ for
every $n\in \NN$. By construction of $V_n\equiv V(x_n, \omega_{n,P_n})$, we have  
$$\dist(0,\partial V_n) \geq\frac{r(P_n)}{2}= s(V_n), \, n \in \NN.$$  
This implies $s(V_n)\longrightarrow \infty$, $n\longrightarrow \infty$. \hfill \qedsymbol  
  
\medskip

Our next aim is to show that a sequence of convex sets with increasing
inradii must be van Hove. We need the following two lemmas.

\begin{lemma} For every $d\in\NN$, there exists a constant $c = c(d)$ with  
$$ (1 + s)^d - (1-s)^d \leq c s$$  
for  $|s|\leq 1$.   
\end{lemma}  
{\it Proof.} This follows by a direct computation.\hfill \qedsymbol  
  
\medskip  
  
For $C\subset \RR^d $ and $\lambda \geq 0$ we set  
  
$$  \lambda C \equiv \{ \lambda x : x\in C\}.$$  
  
\begin{lemma} Let $C$ be a compact convex set in $\RR^d$ with $B(0,s)\subset C$, then the inclusion  
$$C^h \setminus C_h\subset (1 + \frac{h}{s}) C \setminus (1-\frac{h}{s}) |C|$$  
holds, where we set $(1- h s^{-1}) C =\emptyset $ if $h>s$.  In particular,  
  
$$ |C^h \setminus C_h |\leq \kappa \max\{\frac{h}{s} , \frac{h^d}{s^d}\} |C|,$$  
with a suitable constant $\kappa = \kappa(d)$.   
\end{lemma}  
{\it Proof.} The first statement follows by convexity of $C$. The second is then an immediate consequence of the change of variable formula combined with the foregoing lemma. \hfill \qedsymbol

\begin{lemma} \label{convex} Let $(C_n)$ be a sequence of convex sets in $\RR^d$ with $s(C_n) \longrightarrow \infty$, $n\longrightarrow \infty$. Then $(C_n)$ is a van Hove sequence.   
\end{lemma}  
{\it Proof.} Let $h>0$ be given and assume without loss of generality
that $B(0,s (C_n))\subset C_n$. The lemma  follows from the foregoing lemma. \hfill\qedsymbol  
  
\medskip

The following consequence of the foregoing results is a key  ingredient of our proof of   
Theorem \ref{BET}.   
  
\begin{prop}\label{crucialvH} Let $\Oomega$ be  minimal and aperiodic. Let $(P_n)$ be   
a sequence in $\CalP_B(\Omega)$ with $s(P_n)\to \infty$, $n\to \infty$. Let   
$(\omega_n)\subset \Omega$ and $x_n \in \omega_{n,P_n}$  be arbitrary. Then,   
$V(x_n, \omega_{n,P_n})$ is a van Hove sequence.   
\end{prop}  
{\it Proof.} By  Lemma \ref{dist}, aperiodicity of $\OOmega$ together 
with $s(P_n)\to\infty$,   
$n\to \infty$ yields $r(P_n)\to \infty$, $n\to \infty$. Therefore, by
Lemma \ref{VvH}, we have $s(V(x_n, \omega_{n,P_n}))\to \infty$, for $n\to
\infty$.  Now, the statement is immediate   
from  Lemma  \ref{convex}. \hfill \qedsymbol  
  
\medskip  
  
We will also  need an analogue of this proposition for arbitrary (i.e. not necessarily   
aperiodic) DDS.  To obtain this analogue we will need some extra effort.   
  
Let a  minimal DDS $\Oomega$ with periodicity lattice $\CalL$  be given. Let $U=U(\CalL)$ be   
the subspace of $\RR^d$ spanned by the $e_j$, $j=1,\ldots, D(\CalL)$ and let   
$P_U : \RR^d\longrightarrow U$ be the orthogonal projection onto $U$.   
The lattice $\CalL$ induces a grid on $\RR^d$. Namely, we can set   
$$G_0\equiv \{ x\in \RR^d : P_U x = \sum_{j=1}^{D(\CalL)} \lambda_j e_j \;;\mbox{with} \;   
0\leq \lambda_j <1, j=1,\ldots, D(\CalL)\}$$  
and  
$$ G_{(n_1,\ldots, n_{D(\CalL)})} \equiv n_1 e_1 + \ldots + n_{D(\CalL)} e_{D(\CalL)} + G_0,$$  
for $(n_1,\ldots, n_{D(\CalL)})\in \ZZ^{D(\CalL)}$.   
  
We  will now  use coloring of Delone sets to obtain new DDS from $\OOmega$. These new systems   
will essentially be the same sets but equipped with a coloring which ``broadens'' the   
periodicity lattice.   
Coloring has been discussed e.g. in \cite{LS2}.    
  
Let $C$ be a finite set. A Delone set with colorings in $C$ is a subset of $ \RR^d \times C$ such that $p_1(\omega)$ is a Delone set, where $p_1 : \RR^d\times C$   
is the canonical projection $p_1 (x,c)=x$.  When referring to an element $(x,c)$ of a colored   
Delone set we also say that $x$ is colored with $c$. Notions as patterns, pattern classes,   
occurrences, diameter etc. can easily be carried over to colored Delone sets.   
  
 Fix  $\omega\in \Omega$ with $0\in \Omega$.  For every $l\in \NN$, we define a DDS as follows:    
Let $\omega^{(l)} $ be  a Delone set with coloring in $\{0,1\}$ introduced by the following   
rule: $x\in \omega$ is colored with $1$ if and only if  there exists  $(n_1,\ldots, n_{D(\CalL)})  
\in \ZZ^{D(\CalL)}$ with  
$$x \in  G_{(l n_1,\ldots, l n_{D(\CalL)})},$$  
in all other cases $x\in \omega$ is colored with $0$. Set $\Omega^{(l)}\equiv \Omega(\omega^{(l)})  
\equiv \overline{\{ T_t \omega^{(l)} : t\in \RR^d\}}$, where the bar denotes the closure in the   
in the canonical  topology associated to colored Delone   
sets \cite{LS2}. Moreover,  the DDS $(\Omega^{(l)}, T)$ is minimal, as can easily be seen   
considering repetitions of patterns in $\omega^{(l)}$. Also,   $(\Omega^{(l)}, T)$ is uniquely   
ergodic if $\Oomega$ is uniquely ergodic, as follows by considering existence of frequencies in   
$\omega^{(l)}$. The important point about $(\Omega^{(l)}, T)$  is the following lemma.  
  
\begin{lemma}\label{ziel} Let  $\Oomega$ be a minimal DDS and $(\Omega^{(l)}, T)$ for $l\in \NN$   
be constructed as above, then $r(\CalL(\Omega^{(l)}))= l\cdot  r(\CalL (\Omega))$.   
\end{lemma}  
{\it Proof.} This is immediate from the construction. \hfill \qedsymbol  
  
\medskip  
  
Now, we can state the following analog of Proposition \ref{crucialvH}.  
  
\begin{prop}\label{crucialvHstern} Let $\Oomega$ be  minimal and $(\Omega^{(n)},T)$, $n\in \NN$   
constructed as above.  For each $n\in \NN$, choose a pattern class  $P_n\in \CalP (\Omega^{(n)})  
$.  Let $(\omega_n)\subset \Omega^{(n)}$ and $x_n \in \omega_{n,P_n}$  be arbitrary. If   
$s(P_n)\to \infty$, $n\to \infty$, then, $V(x_n, \omega_{n,P_n})$ is a van Hove sequence.   
\end{prop}  
{\it Proof.} By the foregoing lemma and Lemma \ref{dist}, we infer  
$$r(P_n)\longrightarrow \infty, n\longrightarrow \infty.$$  
The statement then follows as in the proof of Proposition
\ref{crucialvH}.\hfill\qedsymbol

\section{The ergodic Theorem}\label{Theergodic}  
In this section we prove Theorem \ref{BET}.  The main idea of the proof
is to combine the geometric decompositions studied in the last section
with the almost additivity of $F$ to reduce the study of $F$ on large
patterns to the study o $F$ on smaller patterns.

\medskip  
  
{\it Proof of Theorem \ref{BET}.} (ii) $\Longrightarrow$(i). For every $Q\in \CalP$ the   
function $P\mapsto \sharp_Q (P)$ is almost additive on $\CalP$. Thus, its average   
$\lim_{n\to \infty} |P_n|^{-1} \sharp_Q (P_n)  $ exists along arbitrary van Hove sequences   
$(P_n)$ in $\CalP$. But this easily implies \eqref{ue} which in turn implies    unique   
ergodicity, as discussed in Sectiuon \ref{Setting}.  
  
\medskip  
  
(i) $\Longrightarrow$ (ii). Let $F : \CalP (\Omega)\longrightarrow \CalB$ be almost additive with error  
function $b$.   
Let $(P_n)$ be a van Hove sequence in $\CalP(\Omega)$. We have to show that   
$\lim_{n\to \infty} |P_n|^{-1} F(P_n)$ exists. As $\CalB$ is a Banach space, it is   
clearly sufficient to show that $(|P_n|^{-1} F(P_n))$ is a Cauchy sequence. To do so we will   
provide  $F^{(k)}$ in $\CalB$   such that   
$$\| |P_n|^{-1} F(P_n) - F^{(k)}\| \;\mbox{ is arbitrarily small for $n$ large and $k$ large.}$$  
To introduce $ F^{(k)}$ we proceed as follows: Fix $\omega \in \Omega$ with $0\in \omega$.  We   
will now first consider the case that $\Oomega$ is aperiodic. The other case can be dealt with   
similarly. We will comment on this at the end of the proof.  Let  $B^{(k)}$ be the ball   
pattern class occurring in $\omega$ around zero with radius $k$ i.e.    
\begin{equation}\label{defbk}  
B^{(k)} \equiv [\omega \wedge B(0,k)].  
\end{equation}

Thus,   
$(B^{(k)})$ is a sequence in $\CalP_B(\Omega)$ with  
$k = s(B^{(k)})\longrightarrow \infty$ for$ k\to\infty$   
and the assumptions of Proposition \ref{crucialvH} are satisfied.

As $\Oomega$ is of finite local complexity, the set   
$\{[B(x,2 R(B^{(k)}))\wedge \omega] : x\in\omega, \omega\in \Omega \;\mbox{with}\;   
[B(x,k)\wedge \omega] = B^{(k)}\}$ is finite. We can thus  enumerate its elements by   
$B_j^{(k)}$, $j=1,\ldots, N(k)$ with suitable $N(k)\in \NN$ and $B_j^{(k)}\in \CalP(\Omega)$.   
Let $C_j^{(k)}\equiv C(B_j^{(k)})$ be the cells associated to $B_j^{(k)}$ according to   
Proposition \ref{crucialcell}. By Proposition \ref{crucialvH},  
$$(*)\;\:\mbox{ $(C_{j_k}^{(l_k)})$ is a  van Hove sequence}$$  
for arbitrary  $(l_k)\subset \NN$ with $l_k \to \infty$, $k\to \infty$,  and $j_k\in \{1,\ldots, N(l_k)\}$.  This will be crucial.   
Denote the frequencies of the $B_j^{(k)}$ by $f(B_j^{(k)})$, i.e.  
\begin{equation}\label{freq}  
f(B_j^{(k)}) =\lim_{n\to \infty} |P_n|^{-1}  \sharp_{B_j^{(k)}} P_n.  
\end{equation}  
Define   
$$ F^{(k)} \equiv \sum_{j=1}^{N(k)} f(B_j^{(k)}) F(C_j^{(k)}).$$  
Choose $\epsilon>0$. We have to show that  
$$\| |P_n|^{-1} F(P_n) - F^{(k)})\|<\epsilon, \;\:\mbox{for $n$  and $k$ large}$$  
(as this will imply that $|P_n|^{-1} F(P_n)$ is a Cauchy sequence). By $(*)$, there   
exists $k(\epsilon)>0$ with   
 \begin{equation}\label{stern} |C_j^{(k)}|^{-1} b(C_j^{(k)}) < \frac{\epsilon}{3}\;\:\mbox{for   
every  $j=1,\ldots, N(k)$}  
\end{equation}  
whenever  $k\ge k(\epsilon)$. (Otherwise, we could  find $(l_k)$ in $\NN$ and  $j_k\in \{1,\ldots, N(l_k)\}$ with $l_k\to \infty, k\to \infty$  such that   
$$ |C_{j_k}^{(l_k)}|^{-1} b(C_{j_k}^{(l_k)})\ge \frac{\epsilon}{3} .$$  
Since  $(C_{j_k}^{(k)})$ is a van Hoove sequence by $(*)$,   
this contradicts property (A4) from Definition \ref{definitionAlAd}.)  
  
\medskip  
  
Let $P\in \CalP$ be an arbitrary pattern class. By minimality of $\OOmega$, we can   
choose $Q=Q(P)\subset \RR^d$ with  $[Q \wedge \omega ]=P$.  
  
\medskip  
  
The idea is now to consider the decomposition of $\omega\wedge Q$ induced by the $B^{(k)}$-decomposition   
of $\omega$. This decomposition of $\omega\wedge Q$  will (up to a boundary term) consist of representatives   
of $C_j^{(k)}$, $j=1,\ldots, N(k)$. For $P=P_n$ with  $n\in \NN$ large, the number of representatives of a   
$C_j^{(k)}$ for $j$ fixed occurring in $Q\wedge \omega$ will essentially be given by $f(C_j^{(k)})|P_n|$.   
Together with the almost-additivity of $F$, this will allow us to relate $F(P_n)$ to $F^{(k)}$ in the   
desired way. Here are the details:

Let $I(P,k)\equiv \{ x\in \omega_{B^{(k)}} : B(x,2 R(B^{(k)})) \subset Q\}$.   
Then, by Lemma \ref{voronoi} and Proposition \ref{crucialcell}  
\begin{equation}\label{zerl}  
 Q\wedge \omega= S\wedge \omega \oplus \bigoplus_{x\in I(P,k)} C(x,\omega,B^{(k)})  
\end{equation}  
with a suitable surface type set $S\subset \RR^d$  with  
  
\begin{equation}\label{bfeature}  
S\subset Q\setminus Q_{4 R(B^{(k)})}.   
\end{equation}

The triangle inequality implies  
\begin{eqnarray*}  
\| \frac{F(P)}{|P|}  - F^{(k)}\| &\leq &   \| \frac{F(P) - F([S\wedge\omega]) - \sum_{ x\in I(P,k)}   
F( [C(x,\omega,B^{(k)})])\|}{|P|  }  \\  
& & + \| \frac{F([S\wedge\omega]) + \sum_{x\in I(P,k)}  F(  [C(x,\omega,B^{(k)})]   )  }{|P|}  - F^{(k)}\| \\  
&& \equiv  D_1 (P,k) + D_2 (P,k).  
\end{eqnarray*}  
  
The terms $D_1 (P,k)$ and $D_2(P,k)$ can be estimated as follows.

By almost additivity of $F$, we have   
  
\begin{eqnarray*}  
D_1(P,k)  
&\leq & \frac{ b([S\wedge \omega]) }{|P|}  + \sum_{x\in I(P,k)}   
\frac{b([C(x,\omega,B^{(k)})])}{|C(x,\omega,B^{(k)})|} \frac{ |C(x,\omega,B^{(k)})|}{|P|}\\  
&\leq& \frac{b(P) + b( [ \oplus_{x\in I(P,k)} C(x,\omega,B^{(k)})    ])}{|P|}\\  
& &\qquad +   
\sup \left\{ \frac{b([C(x,\omega,B^{(k)})])}{|C(x,\omega,B^{(k)})|} \, :\,  x\in I(P,k) \right\}.  
\end{eqnarray*}  
In the last inequality we used (A3).

 Fix $k=k(\epsilon)$ from \eqref{stern} and consider the above estimate for $P=P_n$. Then,  
$$D_ 1 (P_n,k) \leq   
\frac{b(P_n) + b( [ \oplus_{x\in I(P,k)} C(x,\omega,B^{(k)})    ])}{|P_n|} + \frac{\epsilon}{3}.$$  
  
As $(P_n)$ is a van Hove sequence, it is clear  from \eqref{bfeature} that   
$([ \oplus_{x\in I(P,k)} C(x,\omega,B^{(k)})    ] ) $ is a van Hove sequence as well.  Thus   
$$\frac{b(P_n) + b( [ \oplus_{x\in I(P_n,k)} C(x,\omega,B^{(k)} )]}{|P_n|}  = \frac{b(P_n)}{|P_n|}$$  
$$  +   
\frac{   b( [ \oplus_{x\in I(P_n,k)} C(x,\omega,B^{(k)} )])}{ |[ \oplus_{x\in I(P_n,k)}   
C(x,\omega,B^{(k) })]|}  \frac{ |[ \oplus_{x\in I(P_n,k)} C(x,\omega,B^{(k) })]|  }{   |P_n|  }$$  
  tends to zero for $n$ tending to infinity by the definition of $b$. Putting this together, we infer   
$$D_1(P_n,k)<\frac{\epsilon}{2}$$  
for large enough $n\in \NN$.   
  
\medskip  
  
Consider now $D_2$. Invoking the definition of $F^{(k)}$,  we clearly have   
$$  
D_2 (P,k) \leq    
\frac{\|F([S\wedge \omega])\|  }{|P|}   
$$  
$$+ \sum_{j=1}^{N(k)}   
\left|   
\frac{  
\sharp\{ x\in I(P,k) :  [ B(x,2 R(B^k))\wedge \omega ]= B_j^{(k)} \}    
}{|P|} - f(B_j^{(k)})  
\right| \|F(C_j^{(k)})\|.  
$$  
  
Choose $k$ as above and consider $P=P_n$.  By  \eqref{bfeature} and the   
almost additivity of $F$ (property (A2)), we infer that the first term tends to zero for   
$n$ tending to infinity. Again by \eqref{bfeature} and the definition of the frequency, we   
infer that the second term tends to zero as well. Thus,  
$$D_2(P_n,k) < \frac{\epsilon}{2}$$  
for $n$ large. Putting these estimates together, we infer  
$$ \||P_n|^{-1} F(P_n) - F^{(k)}\| \leq D_1 (n,k) + D_2 (n,k) < \epsilon$$  
for large $n$ and the proof is finished for aperiodic DDS.  
  
\medskip  
  
For arbitrary strictly ergodic DDS, we replace the definition of $B^{(k)}$ in \eqref{defbk}, by  
$$ B^{(k)}\equiv [ B(0, k)\wedge \omega^{(k)} ], $$  
where $\omega^{(k)}\in \Omega^{(k)}$ is defined via colouring; see the paragraphs preceding Lemma  
\ref{ziel} in  
 Section \ref{Decomposing}. Then   
$B^{(k)}$ belongs to  $\CalP_B(\Omega^{(k)} )$ for every $k\in \NN$ and    
$$s^k \equiv s(B^{(k)})\longrightarrow \infty, k\longrightarrow \infty.$$   
Thus,  Proposition \ref{crucialvHstern} applies. The proof then proceeds along  the same lines as above, with $\Omega$ replaced by $\Omega^{(k)}$ and  Proposition   
\ref{crucialvH} replaced by Proposition \ref{crucialvHstern} at the corresponding places.  \hfill \qedsymbol  
  
\begin{remark} Using what could be called the \emph{$k$-cells}, $C^{(k)}_j$, $k\in\NN,j=1,\ldots , N(k)$  
from the preceding proof we have actually proven that  
$$  
\lim_{k\to\infty}\sum_{j=1}^{N(k)}f(B^{(k)}_j)F(C^{(k)}_j)=\lim_{n\to\infty}  
\frac{F(P_n)}{|P_n|} .  
$$  
\end{remark}  
{\it Proof of Corollary \ref{lc}.} We use the notation of the corollary.  Apparently, the    
 reasoning  yielding (i) $\Longrightarrow $ (ii) in the foregoing proof remains  valid for   
arbitrary seminorms $\|\cdot\|$. Thus, if $F$ is almost-additive with respect to seminorms   
$\|\cdot\|_\iota$, $\iota\in \CalI$,  then  $(|P_n|^{-1} F(P_n))$, is a Cauchy sequence with   
respect to $\|\cdot\|_\iota$ for every $\iota\in\CalI$. The corollary now follows from   
completeness. \hfill \qedsymbol

\medskip  
{\it Proof of Corollary \ref{mengenversion}.} This can be shown by mimicking the arguments in the above proof. Alternatively, one can define the function   
$\widetilde{F} : \CalP \longrightarrow \CalB$  
by setting $\widetilde{F} (P) := F(Q, \omega)$, where $(Q,\omega)$ is arbitrary with $P = [\omega \wedge Q]$. This definition may seem very arbitrary. However, by (A0), it is not hard to see that $\widetilde{F}(P)$ is (up to a boundary term) actually independent of the actual choice of $Q$ and $\omega$. By the same kind of reasoning, one infers that $\widetilde{F}$ is almost-additive. Now, existence of the limits $|P_n|^{-1} \widetilde{F} (P_n)$ follows for arbitrary van Hove sequnences $(P_n)$. Invoking (A0) once more the corollary follows.  \hfill \qedsymbol  
  
\section{ Uniform convergence of the integrated density of states}\label{Almost}   
This section is devoted to a proof of Theorem \ref{AAT} and Theorem \ref{density}. We need some preparation.  
  
\begin{lemma}   
Let $B$ and $C$ be selfadjoint operators in  a finite dimensional Hilbert space.  Then,   
$|n(B) (E) - n(B+C) (E)|\leq rank(C)$ for every $E\in \RR$, where  $n(D)$ denotes the eigenvalue   
counting function of $D$, i.e. $n(D)(E)\equiv \sharp\{\mbox{Eigenvalues of $D$ not exceeding $E$}\}$.   
\end{lemma}   
{\it Proof.} This is a  consequence of the minmax principle, see e.g. Theorem 4.3.6 in \cite{HoJ}   
for details. \hfill \qedsymbol

\medskip  
  
From this lemma we infer the following proposition.  
  
\begin{prop}\label{subspace} Let $U$ be a subspace of the finite dimensional Hilbert space $X$ with   
inclusion $j : U \longrightarrow X$ and orthogonal projection $p : X\longrightarrow U$.  Then,   
$|n(A) (E) - n(p A j)(E)|\leq  4\cdot rank (1-j\circ p)$ for every selfadjoint operator  $A$ on $X$.   
\end{prop}  
{\it Proof.} Let $P : X \longrightarrow X$ be the orthogonal projection  onto $U$, i.e.   
$P=j \circ p$.  Set $P^\perp\equiv 1-P$ and denote the range of $P^\perp$ by $U^\perp$.  By   
$$ A -P A P = P^\perp A P + P A P^\perp + P^\perp A P^\perp,$$  
and the foregoing lemma, we have $|n(A)(E) - n(PAP)(E)|\leq 3 rank(P^\perp)$. As obviously,   
$$ P A P = p A j \oplus 0_{U^\perp},$$  
with the zero operator $0_{U^\perp} : U^\perp\longrightarrow U^\perp$, $f\mapsto 0$, we also have  
$$ |n(PAP)(E) - n(pAj)(E)|\leq dim (U^\perp).$$  
 As $\dim U^\perp= rank(P^\perp)$, we are done. \hfill \qedsymbol

\medskip  
  
\begin{lemma}\label{combinat} Let $\Oomega$ be an $(r,R)$-system and $\omega\in \Omega$ and $Q$ a   
bounded subset of $\RR^d$. Then, $$\sharp Q\cap \omega \leq \frac{1}{|B(0, r)|} |Q^{r}|.$$  
\end{lemma}  
{\it Proof.} As $\Oomega$ is an $(r,R)$-system, balls with radius $ r$ around different points in   
$\omega$ are disjoint and the lemma follows. \hfill \qedsymbol  
  
\medskip  
  
Our main tool will be the following consequence of the foregoing two results.  
  
\begin{prop}\label{dimension} Let $\Oomega$ be an $(r,R)$-system.  Let   
$Q, Q_j\subset \RR^d$, $j=1,\ldots, n$ be given with $Q=\cup_{j=1}^n Q_j$ and the   
$Q_j$ pairwise disjoint up to their boundaries. Set  
$ \delta(\omega,s) \equiv |\dim \ell^2 (Q_s \cap \omega) - \dim \ell^2 (\cup_{j=1}^n( Q_{j,s} \cap \omega))|$   
for $\omega\in\Omega$ and $s>0$ arbitrary.   
Then,   
\begin{eqnarray*}\delta(\omega,s)&\leq& |\dim \ell^2 (Q \cap \omega) -   
\dim \ell^2 (\cup_{j=1}^n( Q_{j,s} \cap \omega))|\\  
&\leq &  \frac{1}{|B(0,r)|}  \sum_{j=1}^n |Q_j^{r}\setminus Q_{j,s +r}|.  
\end{eqnarray*}  
\end{prop}  
{\it Proof.} Apparently   
$$\dim \ell^2 ( \cup_{j=1}^n( Q_{j,s} \cap \omega)  ) \leq \dim \ell^2 (Q_s \cap \omega)\leq   
\dim \ell^2 (Q \cap \omega).$$  
Now the first inequality is clear and the second follows by   
\begin{eqnarray*}  
\dim \ell^2 (Q \cap \omega) - \dim \ell^2 (\cup_{j=1}^n Q_{j,s} \cap \omega)   
&\leq & \sum_{j=1}^n \sharp ( (Q_j\setminus Q_{j,s}) \cap \omega)\\  
\leq   
\frac{1}{|B(0,r)|}  \sum_{j=1}^n |Q_j^{r}\setminus Q_{j,s +r}|.  
\end{eqnarray*}  
Here, the last inequality follows by the foregoing lemma.  \hfill\qedsymbol   
  
\medskip  
  
We are now able to prove Theorem \ref{AAT}.  
  
\medskip  
  
{\it Proof of Theorem \ref{AAT}.} We have to provide $b :\CalP(\Omega)\longrightarrow (0,\infty)$,   
and  $D>0$ such that (A1) , (A2) and (A3) of Definition \ref{definitionAlAd} are satisfied. Set   
$$ D \equiv \frac{2}{ |B(0,r)|}$$  
and define $b$ by  
$$ b(P) \equiv   \frac{8}{ |B(0,r)| } |Q^{r}\setminus Q_{R^{A} + r} |$$  
whenever $P\in \CalP(\Omega)$ with $P= [(Q,\Lambda)]$. Apparently, $b$
is well defined. Moreover, (A4) follows by the very definition of $b$
and  the  van Hove property.
 
\medskip  
  
Now, (A2) is satisfied as   
$$ \|F^A (P)\| = \|n(A_\omega,Q_{R^A})\|_\infty \leq \sharp ( Q_{R^A}\cap \omega)   
\leq \frac{1}{B(0,r)} |Q^{r}| \leq D |P| + b(P),$$  
for $P=[Q\wedge \omega]$. (A3) can be shown by a similar argument. It remains to show (A1).    
Let $P= \oplus_{j=1}^n P_j$. Then, there exists $\omega\in \Omega$ and bounded measurable sets   
$Q, Q_j$, $j=1,\ldots,n$ in $\RR^d$ with $Q_j$ pairwise disjoint up to their boundaries and   
$Q=\cup_{j=1}^n Q_j$ such that   
$$P = [Q\wedge \omega] \;\:\mbox{and}\:\; P_j =  [Q_j\wedge\omega], j=1,\ldots, n.$$  
As $A$ is an operator of finite range, it follows from the definition of $R^A$ that   
$$ A_\omega|_{\cup_{j=1}^n Q_{j,R^A}} = \oplus_{j=1}^n A_\omega|_{Q_{j,R^A}}$$  
and in particular,  
\begin{equation}\label{senf}  
\sum_{j=1}^n n(A_\omega, Q_{j,R^A}) = n(  A_\omega,\cup_{j=1}^n Q_{j,R^A}).  
\end{equation}  
Thus, we can calculate as follows  
  
\begin{eqnarray*}  
\| F^A (P) - \sum_{j=1}^n F(P_j)\|   
&=& \| n(A_\omega, Q_{R^A}) - \sum_{j=1}^n n (A_\omega, Q_{j,R^A})\|_\infty\\  
&=& \| n(A_\omega, Q_{R^A}) - n(  A_\omega, \cup_{j=1}^n Q_{j,R^A})\|_\infty\\  
(\mbox{Prop \ref{subspace}})\;\: &\leq&  4 (\dim \ell^2 (Q_{R^A}) - \dim \ell^2 (\cup_{j=1}^n Q_{j,R^A})\\  
 (\mbox{Prop \ref{dimension}})\;\: &\leq& \frac{4}{|B(0,r)|} \sum_{j=1}^n | Q_j^{r}   
\setminus Q_{j,R^A + r}|\\  
&\leq& \sum_{j=1}^n b(P_j).  
\end{eqnarray*}  
This finishes the proof. \hfill\qedsymbol

\medskip  
   
We can now proceed to show Theorem \ref{density}. The theorem will be an immediate consequence of   
Theorem \ref{BET} and Theorem \ref{AAT}, once we have proven the  following lemma.  
  
\begin{lemma} Let $\OOmega$ be a strictly ergodic $(r,R)$-system.  Let $A$ be a finite range operator   
with range $R^A$. Then,  $\| n(A_\omega, Q) - F^A ([\omega\wedge Q])\|_\infty \leq 4 |B(0,r)|^{-1}   
|Q^{r}\setminus Q_{R^A + r}|$  
for all $\omega\in \Omega$ and all bounded subsets $Q$ in $\RR^d$.    
\end{lemma}  
{\it Proof} By definition of $F^A$, we have   
$$\| n(A_\omega, Q) - F^A ([\omega\wedge Q])\|_\infty= \| n(A_\omega, Q) -  n(A_\omega, Q_{R^A})\|_\infty.$$  
Invoking Proposition \ref{subspace}, we see that the difference is bounded by $4 \sharp  (Q\setminus   
Q_{R^A})\wedge \omega$. The statement of the lemma now  follows by Lemma \ref{combinat}. \hfill \qedsymbol.   
  
\medskip  
  
{\it Proof of Theorem \ref{density}.}  Let $(Q_n)$ be a van Hove sequence. Then $([Q_n\wedge\omega])$ is a   
van Hove sequence in $\CalP(\Omega)$ independent of $\omega$. Thus, $|Q_n|^{-1} F^A ([Q_n\wedge\omega])$   
converges  uniformly in $\omega\in \Omega$ by Theorem \ref{BET} and Theorem \ref{AAT}. The proof follows   
from the foregoing lemma. \hfill \qedsymbol


\begin{thebibliography}{ABCD}  
  
    
\bibitem{ATF} {\sc M. Arai, T. Tokihiro} and {\sc T. Fujiwara},    
Strictly localized states on a two-dimensional Penrose lattice,    
{\it Phys. Rev. B} {\bf 38} (1988) 1621-1626    
    
      
    
%
\bibitem{BM} {\sc M. Baake} and {\sc R.V. Moody, eds}   Directions in Mathematical   
Quasicrystals, CRM  Monograph series, AMS, Providence RI (2000),  
  
%
\bibitem{be2} {\sc J. Bellissard}, $K$--theory of $C^*$--algebras in solid    
state physics. In: Statistical mechanics and field theory: mathematical     
aspects    
%
%
    
    
    
\bibitem{BHZ} {\sc J. Bellissard}, {\sc  D. J. L. Hermann}, and     
{\sc  M. Zarrouati}, Hulls of Aperiodic Solids and Gap Labelling Theorem,     
In:  Directions in mathematical quasicrystals, CRM Monogr. Ser., 13, Amer.     
Math. Soc., Provicence, RI, 2000,    207--258    
%
  
\bibitem{Cas} {\sc M. Casdagli}, Symbolic dynamics for the renormalization map of a quasiperiodic Schr\"odinger equation, \textit{Commun. Math. Phys.}  {\bf 107} (1986), 295--318  
  
%
%

\bibitem{CS} {\sc W. Craig}, {\sc B. Simon}, Log H\"{o}lder continuity of the integrated density of states for stochastic Jacobi matrices. \textit{Comm. Math. Phys.} \textbf{90} (1983),  207--218

\bibitem{ni} {\sc A. Connes}, Sur la th\'eorie non commutative de     
l'int\'egration. LNM, vol. 725, Springer, Berlin, 1979    
%
\bibitem{Dam} {\sc D. Damanik}, Gordon type arguments in the theory of one-dimensional   
quasicrystals, in \cite{BM}, pp  
\bibitem{del} {\sc B. Delaunay [B.N. Delone]}, Sur la sph\'ere vide,    
{\it Izvestia Akad Nauk SSSR Otdel. Mat. Sov. Nauk.} \textbf{7} (1934),    793-800    
%

\bibitem{DS} {\sc F.  Delyon}{ \sc B. Souillard},Remark on the
    continuity of the density of states of ergodic finite difference
    operators, \textit{Comm. Math. Phys.} \textbf{94} (1984),  289--291

\bibitem{Du1} {\sc F. Durand},  A characterization of substitutive
  sequences using return words,  \textit{ Discrete Math.}  \textbf{179}
  (1998) 89--101  
  
\bibitem{Du2} {\sc F. Durand}, Linearly recurrent subshifts have a
  finite number of non-periodic  subshift factors, {\it Ergodic theory \& Dyn. Syst.}, {\bf 20}, (2000) 1061-1078  
    
\bibitem{FATK} {\sc T. Fujiwara T., M. Arai, T. Tokihiro} and {\sc M. Kohmoto}, Localized states and self-similar states of electrons on a    
two-dimensional Penrose lattice, {\it Phys. Rev. B} {\bf 37}    
(1988) 2797-2804    
    
\bibitem{GH} {\sc C.P.M. Geerse} and {\sc A. Hof}, Lattice gas models    
on self-similar aperiodic tilings, {\it Rev. Math. Phys.}, 3, 1991,     
   163-221      
%
\bibitem{Hof1} {\sc A. Hof}, Some remarks on discrete aperiodic     
Schr\"{o}dinger operators,    
\textit{J.  Statist. Phys.}, 72, (1993)    1353--1374    
%
\bibitem{Hof2} {\sc A. Hof}, A remark on Schr\"{o}dinger operators on     
aperiodic tilings,  \textit{ J. Statist. Phys.}, 81, (1996)    851--855    
%
  
\bibitem{HoJ} {\sc R. Horn} and {\sc C.R. Johnson}, Matrix Analysis, Cambridge University   
Press, Cambridge (1985)  
  
\bibitem{Ni}  
T.~Ishimasa, H.~U.~Nissen and Y.~Fukano,  
\textit{New ordered state between crystalline and amorphous  
in {\sf Ni-Cr}\/ particles},  
Phys.\ Rev.\ Lett.\ {\bf 55} (1985) 511--513.  
  
  
\bibitem{Jan} {\sc C. Janot}, {\it Quasicrystals: A Primer}, Oxford    
University Press, Oxford, 1992    
%
\bibitem{Kel} {\sc J. Kellendonk}, Noncommutative geometry of tilings and gap    labelling, {\it Rev. Math. Phys.}, 7, 1995,    1133-1180    
%
%
\bibitem{KP} {\sc J. Kellendonk} and {\sc I. F.  Putnam}, Tilings;     
$C^\ast$-algebras, and $K$-theory. In: Directions in mathematical     
quasicrystals, CRM Monogr. Ser., 13, Amer. Math. Soc., Provicence,     
RI, 2000,    177-206   
%
\bibitem{KLS} {\sc S. Klassert, D. Lenz} and {\sc P. Stollmann}, Discontinuities  
for the integrated density of states. {\it Comm. Math. Phys.}, to appear  
    
\bibitem{KS} {\sc M. Kohmoto} and {\sc B. Sutherland},    
Electronic States on a Penrose Lattice, {\it Phys. Rev. Lett.}    
{\bf 56} (1986) 2740-2743    
    
    
  
\bibitem{KF} {\sc M. Kraj\v{c}\'{\i}} and {\sc T. Fujiwara},    
Strictly localized eigenstates on a three--dimensional Penrose    
lattice, \textbf{Phys. Rev. B},  {\bf 38} (1988) 12903-12907    
   
    
\bibitem{l1} {\sc J. C. Lagarias}, Geometric Models for Quasicrystals I.    
Delone Sets of Finite Type, \textit{Discrete Comput. Geom.},  \textbf{21} (1999), 161--191. 
%
%
\bibitem{lp} {\sc J. C. Lagarias} and {\sc P.A.B. Pleasants}, Repetitive    
Delone sets and Quasicrystals, {\it Ergodic Theory Dynam. Systems}, to    
appear    
%
\bibitem{LMS} {\sc J.-Y. Lee}, {\sc R.V. Moody} and {\sc B. Solomyak},     
Pure Point Dynamical and Diffraction Spectra, \textit{Ann. Henri
  Poincar\'{e}}, \textbf{3} (2002), 1003--1018. 
%
  
\bibitem{Len2} {\sc D. Lenz}, Uniform ergodic theorems for subshifts over finite alphabets,    
{\it Ergodic Theory Dynam. Systems},     
  
\bibitem{Len3} {\sc D. Lenz}, Hierarchical structures in Sturmian dynamical systems,   
{\it Theoret. Comput. Science}, \textbf{303} (2003), 463--490
  
  
  
%
    
\bibitem{LS1}  {\sc D. Lenz} and {\sc P. Stollmann}, Delone dynamical     
systems, groupoid von Neuman algebras and Hamiltonians for quasicrystals,   
\textit{C. R. Acad. Sci. Paris, Ser. I}, \textbf{334} (2002), 1131 -- 1136  
  
  
\bibitem{LS2}  {\sc D. Lenz} and {\sc P. Stollmann}, Delone dynamical systems and   
associated random operators, in J.-M. Combes et al. (eds) Proc. "Operator Algebra
  
  
  
\bibitem{LS4}  {\sc D. Lenz} and {\sc P. Stollmann}, Algebras of random operators  
associated to Delone dynamical systems, \textit{Math. Phys. Anal. Geom.}, to  
appear   
  
\bibitem{LS5}  {\sc D. Lenz} and {\sc P. Stollmann}, Aperiodic order and quasicrystals:   
spectral properties. \textit{Ann. Henri Poincar\'e}, \textbf{4} (2003), 787 -- 796  
%
%
\bibitem{Pat} {\sc J. Patera (ed)}, Quasicrystals and Discrete Geometry,  
Fields Institute Monographs, vol. 10, AMS, Providence, RI 1998  
%
  
\bibitem{Pri} {\sc N. Priebe},  Towards a characterization of
  self-similar tilings in terms of dereived Voronoi tesselations, {\it  Geometriae Dedicata}, \textbf{79} (2000) 239--265  
  
\bibitem{Sut} A.\ S\"ut\H{o}, The spectrum of a quasiperiodic Schr\"odinger operator, \textit{Commun.\ Math.\ Phys.} {\bf 111} (1987), 409--415  
  
  
%
  
%
\bibitem{Sen} {\sc M. Senechal}, {\it Quasicrystals and Geometry},    
Cambridge University Press, Cambridge, 1995    
%
\bibitem{She} {\sc D. Shechtman, I. Blech, D. Gratias} and {\sc J.W. Cahn}:    
Metallic phase with long-range orientational order and no translation    
symmetry, {\it Phys. Rev. Lett.}, 53, 1984,    1951-1953     
%
%
\bibitem{Sol1} {\sc B. Solomyak}, Dynamics of self-similar tilings,    
{\it Ergodic Theory Dynam. Systems}, 17, 1997,    695-738    
%
%
  
\end{thebibliography}
\end{document}